\documentclass[prl,superscriptaddress,amsmath,amssymb,twocolumn]{revtex4-2}
\usepackage{graphicx}
\usepackage{subfigure}
\usepackage{adjustbox}
\usepackage{bm}
\usepackage{placeins}
\usepackage{bbm}
\usepackage{color}
\usepackage[dvipsnames]{xcolor} 
\usepackage{braket}
\usepackage{standalone}
\usepackage{multirow}
\usepackage{tikz}
\usepackage{mathrsfs}
\usepackage{dsfont}
\usepackage[colorlinks,bookmarks=true,citecolor=blue,linkcolor=blue,urlcolor=blue]{hyperref}
\usepackage{cleveref}
\usepackage{comment}
\usepackage{tabularx}
\usepackage{mathtools}
\usepackage[sectionbib]{bibunits}
\defaultbibliographystyle{unsrt} 
\defaultbibliography{refs.bib}

\Crefname{equation}{Eq.}{Eqs.}
\Crefname{figure}{Fig.}{Figs.}

\begin{document}




\title{Liouvillian skin effects and fragmented condensates in an integrable dissipative Bose-Hubbard model}




\author{Christopher Ekman}
\email{christopher.ekman@fysik.su.se}
\affiliation{Department of Physics, Stockholm University, AlbaNova University Center, 10691 Stockholm, Sweden}
\author{Emil J. Bergholtz}
\email{emil.bergholtz@fysik.su.se}
\affiliation{Department of Physics, Stockholm University, AlbaNova University Center, 10691 Stockholm, Sweden}

\date{\today}

\begin{abstract}
Strongly interacting non-equilibrium systems are of great fundamental interest, yet their inherent complexity make then notoriously hard to analyze. We demonstrate that the dynamics of the Bose-Hubbard model, which by itself evades solvability, can be solved exactly at any interaction strength in the presence of loss tuned to a rate matching the hopping amplitude. Remarkably, the full solvability of the corresponding Liouvillian, and the integrability of the pertinent effective non-Hermitian Hamiltonian, survives the addition of disorder and generic boundary conditions. By analyzing the Bethe ansatz solutions we find that even weak interactions change the qualitative features of the system, leading to an intricate dynamical phase diagram featuring non-Hermitian Mott-skin effects, disorder induced localization, highly degenerate exceptional points, and a Bose glass-like phase of fragmented condensates. We discuss realistic implementations of this model with cold atoms.  

\end{abstract}
\maketitle

{\it Introduction.--}
In cold atom systems with strong interactions \cite{RevModPhys.80.885,Kinoshita2006,Lewenstein2007,Gross2017,Zhang2018} and disorder \cite{Garreau2017}, dissipation is an ever-present adversary, since environmental coupling tends to destroy intriguing, but delicate, quantum phenomena. However, it is being increasingly appreciated that dissipation itself may in fact be harnessed \cite{diehl2008quantum,Bardyn_2013,goldman2016topological,Müller2021, Helmrich2020,Sponselee_2019,Gou_2020,Liang_2022,Bouchoule2021,Yamamoto2023,mengdissipative,fandissipative,Gong2018,ashida2020non,bergholtzreview} and induce unique intriguing effects \cite{Gong2018,ashida2020non,bergholtzreview,grun2022protocol}, thus providing a new frontier in the study of non-equilibrium quantum many-body systems \cite{eisert2015quantum}. The vast majority of recent studies have focused on non-interacting dissipative systems  \cite{Gong2018,ashida2020non,bergholtzreview}. In the quantum realm these systems may realize novel topological phenomena such as the Liouvillian version \cite{Song2019,PhysRevLett.127.070402,PhysRevResearch.4.023160, Mao_2024} of the non-Hermitian skin effect \cite{lee2016,yao2018,kunst2018,martinez2018,Okuma2020,Okuma2023,lin2023} in which a macroscopic number of eigenstates are exponentially localized at the boundaries of the system. Several recent studies investigate the interplay between interactions and the skin effect \cite{Yoshida2023,Kim2023,Yang2021,Mao_2023,Wang_2023,zheng2023exact,fate22,Shen_2022,hamanaka2024multifractality,zhong2024density}, yet this topic remains far less explored than the non-interacting case. 

Even in absence of dissipation, strongly interacting models are difficult to analyze. This in particular applies to the paradigmatic one-dimensional Bose-Hubbard model \cite{bosehubbard,bosehubbardtool}, which is famously non-integrable \cite{Choy1982} for more than two sites \cite{Links2003}, in contrast to its fermionic counterpart. The absence of integrability combined with its key experimental importance \cite{Jaksch1998} has made it a popular target in numerical studies, using, for example density, renormalization group techniques \cite{Kollath_2004,Kuehner_1998,Ejima_2011,Urba_2006,Schmidt_2007,kollath2007}, Monte Carlo methods \cite{Kashurnikov_1996,Pollet_2013}, and exact diagonalization \cite{Sowinski_2012,Zhang_2010,kollath2007}.

A seminal exact solution strategy of a strongly interacting many body system was provided by Bethe \cite{Bethe} in his famous exact solution of the Heisenberg model. Much more recently, third quantization \cite{Prosen_2008} was invented as a framework to solve quadratic (non-interacting) Liouvillians, describing driven-dissipative systems, and in the last few years the notion of integrability was extended to more general systems including certain interacting Liouvillians \cite{Medvedyeva_2016, de_Leeuw_2021,Ziolkowska_2020, Buča2020,Nakagawa2021}. The application of integrability methods to the solution of Liovillians has thus made it possible to construct exactly solvable models that combine both dissipation and strong interactions, but exactly solvable models of that kind are rare, and exactly solvable models that incorporate disorder, dissipation and interactions simultaneously remain elusive.

\begin{figure}
    \includegraphics[width=0.5\textwidth] {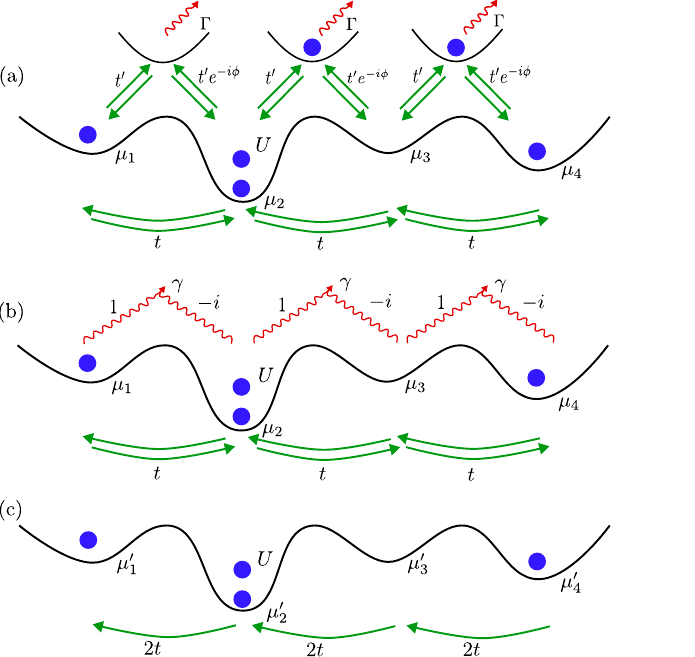} 
    \caption{The models under consideration in this letter. (a) A triangular ladder as implemented in Refs. \onlinecite{Gou_2020} and \onlinecite{Liang_2022}, with an additional on-site potential and interactions of strength $U$. The upper sites have dissipation with rate $\Gamma$, while the lower ones do not. A fictitious magnetic flux $\phi$ leads to a relative phase in the hopping terms between the different sites. Upon integrating out the green sites we obtain (b) with dissipation acting on neighbouring sites, where the magnetic flux has been chosen to be $\phi=\frac{\pi}{2}$. (c) Shows the system described by the integrable unidirectional effective non-Hermitian \eqref{NH_hamiltonian}, where the dissipation has been absorbed into the hopping parameter and the on-site potential.}\label{main_fig}
\end{figure}

In this Letter, we present and analytically solve a disordered Bose-Hubbard chain with environmental coupling that is tailored to render it exactly solvable while remaining realistic in cold atom systems (see Fig. \ref{main_fig}). In the process we derive and solve the non-Hermitian Hamiltonian of the unidirectional Bose-Hubbard chain that was first shown to be Yang-Baxter integrable in Ref. \onlinecite{zheng2023exact}. Here we extend this analysis in several ways that reveal qualitatively new physics. We show how the model emerges as an effective description of a realistic setup, and that it remains integrable in the presence of arbitrary on-site potentials, including random on-site disorder, and with open boundary conditions. The open boundary conditions enable non-Hermitian Mott-skin effects and highly degenerate exceptional points, and adding disorder leads to a novel phase that is reminiscent of a Bose glass. Furthermore, we stress that our analysis goes beyond the effective non-Hermitian Hamiltonian approach by providing a complete solution of the Lindblad Liouvillian.


{\it Model and physical realization.--}
Dissipative cold atom systems may be realistically modeled using the Lindblad master equation \cite{Lindblad,Breuer},  
\begin{equation} \label{Lindbladian}
    \frac{d\rho}{dt}=\mathcal{L}\rho=-i[H,\rho]+\sum_j  \gamma_j\left(2\ell_j\rho\ell^{\dagger}_j-\{\ell_j\ell_j^{\dagger},\rho\}\right).
\end{equation}
with $\gamma_j\geq0$.
We consider the coherent dynamics to be described by the disordered Bose-Hubbard Hamiltonian
\begin{align} 
    H=-&t\sum_{j=1}^N a_j^{\dagger}a_{j+1}+a_{j+1}^{\dagger}a_j+\frac{U}{2}\sum_{j=1}^N n_j(n_j-1)+\sum_{j=1}^N  \mu_j n_j,\nonumber\\
    &-\epsilon (a^{\dagger}_{N}a_1+a^{\dagger}_1a_N)\label{bosehubard}
\end{align}
where $a_j$ are bosonic annihilation operators, $n_j=a^{\dagger}_ja_j$ are the number operators, and $\mu_j$ are disorder variables.  

One way to incorporate realistic dissipation so that the system becomes exactly solvable is to consider a triangular ladder \cite{Liang_2022,Gou_2020}, as in Fig. \ref{main_fig}(a). Particles dissipate from the sites on the top and these sites can be integrated out \cite{supplemental_material} (see also Ref. \onlinecite{Kamenevbook}), leaving a chain with jump operators  \begin{equation} \ell_j=a_j-ia_{j+1}\label{jump}\end{equation} as in Fig. \ref{main_fig}(b), cf. Ref. \onlinecite{Song2019}. The relative phase factor may be obtained experimentally by including a magnetic field. Remarkably, at the level of the effective non-Hermitian Hamiltonian, the hopping in one direction vanishes if $\gamma_j=t$ for $j\neq N$ and $\gamma_N=\frac{\epsilon}{2}$. Explicitly, the resulting effective non-Hermitian Hamiltonian (Fig. \ref{main_fig}(c)) becomes
\begin{align} \label{NH_hamiltonian}
    H_{eff}\!&=\!-2t\sum_{j=1}^N  a_{j}^{\dagger}a_{j+1}\!+\!\frac{U}{2}\sum_{j=1}^N n_j(n_j-1)\!+\!\sum_{j=1}^N  \mu_j'n_j\!-\!\epsilon a^{\dagger}_1 a_N.
\end{align}
where we defined $\mu'_j=\mu_j-2it$ for $j\neq 1,N$ and $\mu'_1=\mu_1-i(t+\frac{\epsilon}{2}),\mu'_N=\mu_N-i(t+\frac{\epsilon}{2})$. 
Ref. \onlinecite{Liang_2022} realized and analyzed the non-interacting and non-disordered limit of Eq. (\ref{NH_hamiltonian}) in a cold atom system. 
Here we include both interactions and disorder, and in addition to Eq. (\ref{NH_hamiltonian}), also consider the full Liovillian quantum dynamics as described by Eqs. (\ref{Lindbladian}), (\ref{bosehubard}) and (\ref{jump}). An immediate conclusion that can be drawn is that in the absence of interactions a state with $N$ particles decays at a rate of $-\mathrm{Im}\left(\sum_j \mu'_j n_j\right)\approx 2tN$.

{\it Bethe ansatz solution.--} Using the machinery of the quantum inverse scattering method \cite{Korepinbook} it may be shown that the right eigenvectors of the Hamiltonian, and of all its constants of motion defined in the supplemental material \cite{supplemental_material}, have the form given by the algebraic Bethe Ansatz
\begin{equation} \label{ABA}
    \ket{\{\lambda_n\}}=\prod_{n=1}^M \mathcal{B}(\lambda_n)\ket{\Omega}
\end{equation}
for an operator $\mathcal{B}(\lambda)$ that takes a continuous argument $\lambda$. Each application of $\mathcal{B}$ creates a quasiparticle, so the state given above has $M$ quasiparticles. In the Fock basis the vacuum $\ket{\Omega}$ is equal to the state without any particles in it. The states given by Eq.~$\eqref{ABA}$ are eigenvectors of the Hamiltonian if the rapidities $\lambda_n$ solve the Bethe equations, which in this case are given by
\begin{align} \label{Bethe_eqs}
    \prod_{k=1}^N(\lambda_n-\mu'_k)&\prod_{j\neq n}(\lambda_n-\lambda_j-U)\nonumber\\&-\epsilon \left(-\frac{2t}{U}\right)^N\prod_{j\neq n}(\lambda_j-\lambda_n-U)=0.
\end{align}
 In terms of the rapidities the eigenvalue corresponding to the eigenstate \eqref{ABA} of the non-Hermitian effective Hamiltonian is $E(\{\lambda_n\})=\sum_{n=1}^M\lambda_n$.  In the case without an onsite potential, meaning $\mu_k'=0$ for all $k$, and with $\epsilon\neq 0,$ this was first shown in Ref. \onlinecite{zheng2023exact}.


The coordinate Bethe Ansatz in terms of the rapidities is \cite{Jiang_2020}
\begin{equation} \label{coord_bethe}
    \Psi\left(\{x_k\}_{k=1}^M\right)=\sum_{\sigma\in S_M}\prod_{j>k}f(\lambda_{\sigma_j}-\lambda_{\sigma_k})\prod_{n=1}^M\prod_{j=1}^{x_n}\left(\frac{\lambda_{\sigma_n}-\mu_{j}}{-2  t}\right)
\end{equation}
where $f(\lambda)=1-\frac{U}{\lambda}$ and $S_M$ is the set of all permutations of the integers between $1$ and $M$. This expression matches the algebraic Bethe Ansatz. 

{\it Integrability and solvability.--}
The integrability structure outlined above can be used to explicitly diagonalize the Liouvillian \eqref{Lindbladian}. It is important that there is only dissipation (or only gain), and that the effective non-Hermitian Hamiltonian commutes with the number operator, as this ensures that the Liouvillian is triangular in the basis given by states of the form $\ket{\{\lambda_n\}_{n=1}^M}\bra{\{\nu_m\}_{m=1}^{M'}}$ \cite{Torres2014}. Its eigenvalues are given by 
\begin{equation}
    \omega(\{\lambda_n\}_{n=1}^M,\{\nu_m\}_{m=1}^{M'})=-i\sum_{n=1}^M\lambda_n+i\sum_{m=1}^{M'}\bar{\nu}_m
\end{equation}
and the eigenoperators have the form
\begin{equation} \label{Lindblad_eig}
\rho_{(\lambda,\nu)}=\sum_{n=0}^m\sum_{(\lambda',\nu')}v^{(l,m;n)}_{(\lambda,\nu);(\lambda',\nu')}\ket{\{\lambda'\}}\bra{\{\nu'\}},
\end{equation}
where the coefficients are given in the supplemental material \cite{supplemental_material} (see also Refs. \onlinecite{Slavnov1989,Gaudin,Oota2004,Brody2014,Slavnovbook}). Note that the unique steady state of the system is the empty state, so there cannot be any conserved quantities. The Liouvillian therefore cannot easily be said to be integrable, in spite of the underlying integrability structure that allows its exact analytical solution.

\begin{figure}
    \centering
    \includegraphics[width=0.45\textwidth]{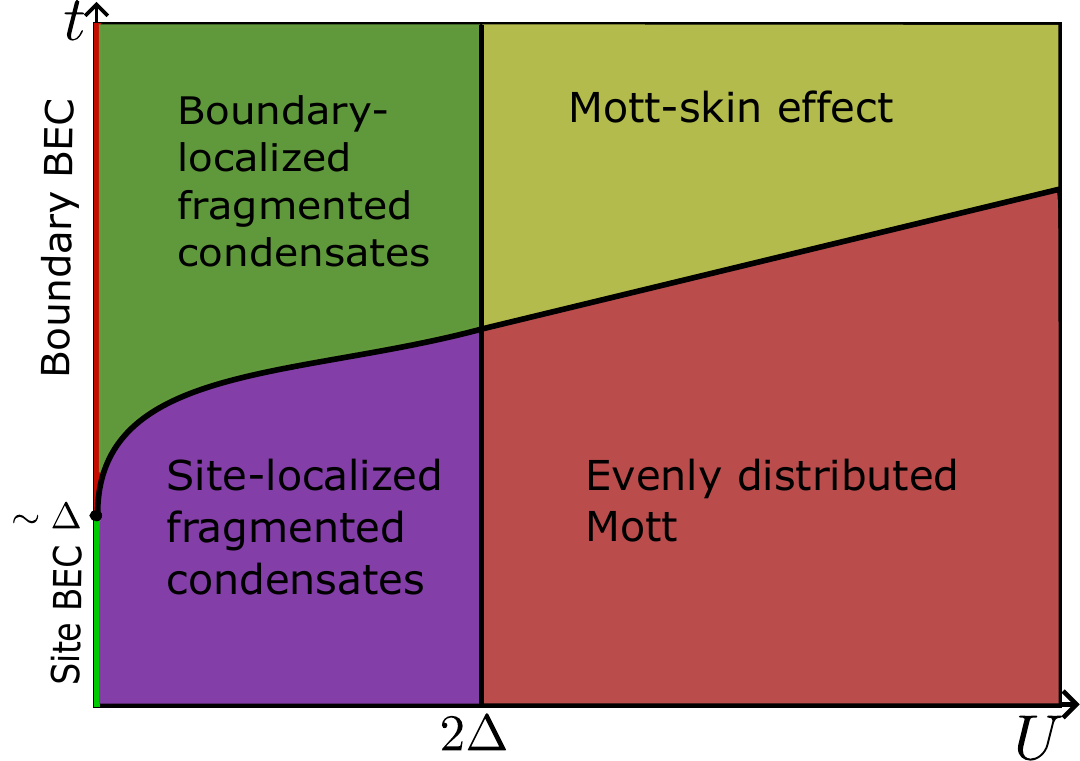}
    \caption{ The ground state phase diagram of the non-Hermitian Hamiltonian with open boundary conditions at integer filling as obtained analytically from the two-point correlation functions in Eqs. \eqref{BG_corr} and \eqref{Mott_corr}. For $U>2\Delta$ the ground state is in a Mott phase, which can either be localized at the boundary or be evenly distributed depending on $t$, whereas for $U<2\Delta$ the system is in Bose-Glass-like phase of fragmented condensates, which may also be localized either at the boundary or at some site. At $U=0$ there is a transition between a Bose-Einstein condensate at the boundary and a Bose-Einstein condensate at a single site in the bulk.}
    \label{fig:GS_phase_diagram}
\end{figure}

{\it Correlation functions and phase diagram for open boundary conditions.--}
We now examine the properties of the model defined by the effective Hamiltonian \eqref{NH_hamiltonian} with random disorder and open boundary conditions ($\epsilon=0$), which exhibits a remarkable degree of solvability. As we detail below this leads to the phase diagram in Fig. \ref{fig:GS_phase_diagram}. For arbitrary disorder the Bethe equations now take the factorized form
\begin{equation}
    \prod_{k=1}^N(\lambda_n-\mu_k')\prod_{j\neq n}(\lambda_n-\lambda_j-U)=0
\end{equation}
which have solutions $\lambda_n^k=\mu_k'+(n-1)U$ where $n=1,2,\ldots$. Inserting this into the expression for the energy we see that the spectrum is given by
\begin{equation} \label{open_energy}
    E(n_1,n_2,\ldots,n_N)=\sum_{k=1}^N \mu_k' n_k+\frac{U}{2}n_k(n_k-1).
\end{equation}
The effective Hamiltonian exhibits exceptional points, i.e. at a non-Hermitian degeneracy at which both eigenvalues and eigenstates coalesce,  whenever there are distinct partitions of integers, $\{n_i
 \}$, giving identical energies in Eq. (\ref{open_energy}). In particular this happens when any two (or more) disorder variables, $\mu_i$, are equal. Moreover, if the potential is constant (no disorder), there are exceptional points of very high order. It is worth noting that although at $t= 0$ the integers $\{n_i
 \}$ labeling energies and eigenstates correspond to particle densities, their interpretation is less straightforward at $t\neq 0$.

In the limit $U\rightarrow 0$ it is sufficient to treat the single-particle eigenstates. These are given by
\begin{equation}
    \Psi_{\lambda}(x)=\prod_{j=1}^x\left(\frac{\lambda-\mu_j'}{-2t}\right)
\end{equation} 
which are eigenstates if $\lambda=\mu_n'$ for $n=1,2,\ldots N$. If there is no disorder, meaning $\mu_n'=0$ for all $n,$ all the eigenstates coalesce to $\psi(x)=\delta_{x,1}$. One simple observation is in that if $\lambda=\mu_n'$ then the particle density is $0$ for $x>m$ where $m$ is the smallest integer such that $\mu_m'=\mu_n'$. The particle is therefore sharply localized to a finite region. In addition, if $t>>|\mu_{m'}'-\mu_{n'}'|$ for all $m',n'$ we see that the particle is exponentially localized in the left side of that region for all eigenstates. On the other hand, if $t<<|\mu_{m'}'-\mu_{n'}'|$ on average each eigenstate is exponentially localized to its corresponding site. Extending to the many-body ground state, we see that it is a gapless condensate of particles in the state with $\lambda$ equal to the $\mu_n'$ with smallest real part.  Due to the simplicity of this state, it is possible to place strong bounds on the two-point correlation function. Let $x_2>x_1$ both be in the region where the wave function is non-zero. Note that the imaginary parts of the $\mu_n$:s cancel in the eigenstates, except for the factors that include $\mu_1'$ and $\mu_N'$. These do not have a qualitative effect on the physics, so we neglect them. Using the notation $\mu_k=\mu_k'+2it$
\begin{align}
    G(x_1,x_2)&=\frac{\bra{GS}a^{\dagger}_{x_1}a_{x_2}\ket{GS}}{\langle GS|GS\rangle}\!\propto
    \prod_{j=1}^{x_1}\!\left(\frac{\mu_n\!-\!\mu_j}{-2t}\right)\!\prod_{j=1}^{x_2}\left(\frac{\mu_n\!-\!\mu_j}{-2t}\right).
\end{align}
In order to proceed we must place some constraints on the disorder variables. Suppose that they are bounded and not all the same. Then we may choose the smallest interval $\mathcal{I}$ such that $\mu_j\in \mathcal{I}$ for all $j$ to be $\mathcal{I}=[-\Delta,\Delta]$ with $\Delta>0$. Clearly $\mu_n=-\Delta$. Suppose further that the least $\mu_{m}>-\Delta$ is given by $\mu_m=-\Delta+\delta$. Then the two-point correlation function satisfies the bounds

\begin{align}
\frac{1}{\langle GS|GS\rangle}\left(\frac{\delta}{2t}\right)^{x_1+x_2}\leq |G(x_1,x_2)|\leq \frac{1}{\langle GS|GS\rangle}\left(\frac{\Delta}{t}\right)^{x_1+x_2}
\end{align}
 for $x_1,x_2<n$. If either $x_1$ or $x_2$ is greater than or equal to $n$ then $G(x_1,x_2)=0$.
We see that the correlation length satisfies $1/\xi\in\!\left[\ln \frac{\delta}{2t},\ln \frac{\Delta}{t}\right]$ and that $\xi$ diverges for some $t\leq\Delta$ signifying a phase transition. One of the phases is a skin-effect-phase, where all the particles are localized at one of the edges (corresponding to the red line at the left edge of Fig. \ref{fig:GS_phase_diagram}), and the other phase is a disorder-induced localized phase, where all the particles are localized at some site (corresponding to the green line at the left edge of Fig. \ref{fig:GS_phase_diagram}). Other states show a similar behaviour. 

 Now we let $U>0$. This has dramatic effects for arbitrarily weak interactions in the thermodynamic limit, $N,M\rightarrow \infty$. In the ground state the particles no longer condense into the same mode, due to the repulsive interactions. Instead, the particles spread out over the different modes, so that there are competing condensates. The Bose-Einstein condensate has fragmented due to the interactions. This is reminiscent of a Bose-Glass \cite{Fisher1989,Scalettar1991}, where the competition between disorder and interactions causes disjoint localized pockets of superfluid condensate to form. Here, in contrast, the condensates are not spatially disjoint unless the disorder is much larger than the hopping parameter.

If $U$ is sufficiently small one of the modes will still have the largest fraction of particles, say $M_0>M/N$ of them, so that there are rapidities $\lambda_{n}=-\Delta+(n-1)U$ for $n=1,2,\ldots M_0$. The leading part of the two-point correlator can then be shown to satisfy
\begin{align} \label{BG_corr}
    D\!\left(\frac{\!(M_0\!-\!1)U\!+\!\delta}{2t}\right)^{X}\!\leq &|G(X)|\leq\! D\!\left(\frac{\!(M_0\!-\!1)U\!+\!2\Delta}{2t}\right)^{X}
\end{align}
where we defined $X=x_1+x_2$ and $D>0$. We see that the correlation length satisfies \linebreak$1/\xi\in\! \left[\ln \frac{(M_0-1)U+\delta}{2t},\ln \frac{(M_0-1)U+2\Delta}{2t}\right]$. 
Clearly $\xi$ diverges at some point, marking a transition between a boundary-localized skin-effect phase (the purple part of Fig. \ref{fig:GS_phase_diagram}) and a site-localized phase (the green part of Fig. \ref{fig:GS_phase_diagram}) where $M_0$ particles are localized at the site where the disorder potential is minimized, while the rest are distributed in condensates elsewhere in the lattice.

If $U>2\Delta$ the energy is minimized by equally distributing the integers in Eq.~\eqref{open_energy}. Suppose for simplicity that $M/N=n$ is an integer. In that case the gap is $U-2\Delta$, so this ground state is a Mott insulator at integer filling. The leading order behaviour of the two-point correlator satisfies
\begin{align} \label{Mott_corr}
    D\left(\frac{(n-1)U+\delta}{2t}\right)^{X}\!\leq &|G(X)|\!\leq D\left(\frac{(n-1)U+2\Delta}{2t}\right)^{X}
\end{align}
again for $D>0$. The correlation length  satisfies $1/\xi\in\! \left[\ln \frac{(n-1)U+\delta}{2t},\ln \frac{(n-1)U+2\Delta}{2t}\right]$
and diverges at some point, signaling a phase transition between a boundary-localized Mott-skin effect phase and a Mott phase where the particles are evenly distributed across the lattice, as shown in the yellow and red parts of Fig. \ref{fig:GS_phase_diagram}, respectively. 

Numerical calculations corroborating these results are presented in the supplemental material \cite{supplemental_material}.

{\it Liouvillian dynamics.-- } Now let us turn to the dynamics of the full Lindblad master equation. For a generic initial state we expect that the time evolution includes every eigenoperator \eqref{Lindblad_eig}, hence the two-point correlator consists of some time-dependent linear combination of all two-point correlators of the model defined by the Hamiltonian \eqref{NH_hamiltonian}. By the same reasoning as for the ground state correlators the leading order time dependent asymptotics at time $\tau$ have the form $G(X;\tau)=\sum_{n=1}^M D_n(\tau)g_n(X)$ where
\begin{equation}
    \left(\frac{(n-1)U+\delta}{2t}\right)^{X}\!\leq |g_n(X)|\!\leq \left(\frac{(n-1)U+2\Delta}{2t}\right)^{X}
\end{equation}
and the $D_n(\tau)$ are obtained from the coefficients in Eq. \eqref{Lindblad_eig}. Since there is only dissipation it follows that the functions $D_n(\tau)\rightarrow 0$ as $\tau\rightarrow \infty$. We see that if $U$ is non-zero there will only be a phase transition between boundary localization and site localization once the mean particle number has decreased to a finite number. If $U=0,$ on the other hand, there is always such a dynamic phase transition.

{\it Discussion.--}
In this work, we have shown that the paradigmatic example of a non-integrable model, namely the Bose-Hubbard model, remarkably becomes exactly Bethe Ansatz solvable in the presence of dissipation terms that may be realistically implemented with cold atoms. The effective non-Hermitian Hamiltonian as well as the full non-equilibrium quantum dynamics feature a rich phase diagram that invites ample future theoretical as well as experimental work.  

As mentioned above, the realization of a momentum-space lattice version of this model in the non-interacting limit in a cold atom setup was reported in Ref.~\onlinecite{Gou_2020}. There the parameters were chosen so that interactions were unimportant and attractive. We expect that similar setups can be tuned to the regime of repulsive interactions described in this paper, although the real-space interactions must be chosen to be attractive in order the be repulsive in the momentum-space lattice \cite{An_2018}. While previous realizations of the non-interacting limit have used $^{87}\mathrm{Rb}$ atoms, $^{39}\mathrm{K}$ may be more suitable for the interacting system, as its relevant Feschbach resonances are wider, allowing for more precise control of the interactions \cite{Meierthesis}. The existing experimental implementations of the non-interacting limit can therefore be modified to include interactions, which we expect would exhibit the phase transitions described above.

The model studied in this work may be generalized to higher dimensions and more involved interaction terms. To this end one has to devise Lindblad operators such that the particles in the system can only hop in one direction along each dimension. Then the Fock basis can be ordered so that the kinetic term is triangular, and consequently the non-Hermitian effective Hamiltonian and the Liouvillian are triangular as long as the remaining terms are polynomial in the on-site densities. If open boundary conditions are chosen it is straightforward to calculate the eigenvalues of the effective Hamiltonian, while determining the corresponding eigenvectors is technically more complicated. 

An interesting conceptual aspect of this work is that the Lindbladian does not have any conserved quantities, yet it is solvable by the Bethe Ansatz, similarly to the models discussed in \cite{Buča2020,Nakagawa2021}. There, however, the system is integrable in the absence of dissipation in contrast to the present case \cite{Choy1982}. 

The topological nature of the non-interacting non-Hermitian skin effect suggests that the quantum many-body generalizations thereof studied here is also robust to generic perturbations, which is promising in the context of its realization.

Finally, we also note that an alternative interpretation of the non-Hermitian Hamiltonian Eq. ~\eqref{NH_hamiltonian} can be gleaned from \cite{Hatano1996,Hatano1997,Lehrer1998}. There the difference in the hopping to the left and to the right, respectively, are obtained from the strength of a magnetic field applied to interacting flux lines in an array of defects in a superconductor. The unidirectional Hamiltonian is obtained when the transverse field component is large. 

\acknowledgments{{\it Acknowledgements.--} We acknowledge helpful discussions with Rodrigo Arouca de Albuquerque and Alexander Fagerlund.
This work was supported by the Swedish Research Council (VR, grant 2018-00313), the Wallenberg Academy Fellows program of the Knut and Alice Wallenberg Foundation (2018.0460) and the G\"oran Gustafsson Foundation for Research in Natural Sciences and Medicine.}

\bibliography{refs.bib}

\pagebreak
\newpage

\onecolumngrid
\pagebreak
\newpage
 		\renewcommand{\theequation}{S\arabic{equation}}
		\setcounter{equation}{0}
		\renewcommand{\thefigure}{S\arabic{figure}}
		\setcounter{figure}{0}
		\renewcommand{\thetable}{S\arabic{table}}
		\setcounter{table}{0}
 		\section{Supplemental Material for 'Liouvillian skin effects and fragmented condensates in an integrable dissipative Bose-Hubbard model'}

    The supplemental material provides derivations and details in addition to the main text. In section 1 the effective Lindbladian used in the main text is derived from a more realistic model. In section 2 the structure underlying the exact solvability of the model is shown, and in section 3 more explicit expression for the eigenoperators of the Liouvillian are derived.
\FloatBarrier

\renewcommand{\theequation}{S\arabic{equation}}
\setcounter{equation}{0}
\renewcommand{\thefigure}{S\arabic{figure}}
\setcounter{figure}{0}
\renewcommand{\thetable}{S\arabic{table}}
\setcounter{table}{0}
\section{1. Derivation of the effective Lindbladian}
\begin{figure}
    \centering
    \includegraphics{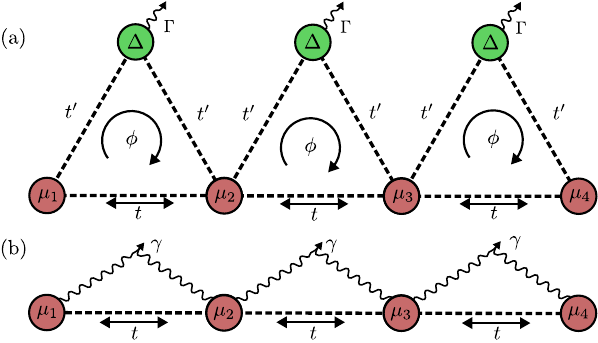}
    \caption{Figure (a) shows the triangular chain, where there is dissipation only in the upper sites. Figure (b) shows the remaining sites after the upper sites have been integrated out. An effective non-local dissipation with decay rate $\gamma$ has emerged, while the on-site potentials and the hopping parameter are renormalized.}
    \label{fig:sm1}
\end{figure}
An effective model describing the cold atom setup shown in figure \ref{fig:sm1} is given by the Lindblad equation
\begin{equation} \label{S_Linblad}
     \frac{d\rho}{dt}=\mathcal{L}\rho=-i[H,\rho]+\sum_j  \Gamma_j\left(2L_j\rho L^{\dagger}_j-\{L_jL_j^{\dagger},\rho\}\right).
\end{equation}
The physics described by this equation may equivalently be captured by a path integral in the general Keldysh formalism, see Ref. \cite{Kamenevbook} for a textbook treatment. 
In this framework the field operators are doubled, with one set living on the forward part of the closed time contour, and the other set living on the backwards path.
Let the annihilation operators acting on the lower sites be given by $a_j$ and the annihilation operators acting on the upper sites be given by $b_j$ for sites $j=1,2,...,N$. In addition we use a superscript $+(-)$ to indicate that a field lives on the forward(backward) part of the time contour. Taking the system to be noninteracting for simplicity it may be described by the following path integral
\begin{equation}
    Z=\int \mathcal{D}\left[\bar{a},a\right]\mathcal{D}\left[\bar{b},b\right]e^{iS[\bar{a},a,\bar{b},b]}
\end{equation}
where the Keldysh action $S$ is given by
\begin{align} \label{keldysh_action}
    S[\bar{a},a,\bar{b},b]=\int dt\  &\sum_{n=1}^N\left(i\bar{b}^{+}_n\frac{d}{dt}b^+_n-i\bar{b}^{-}_n\frac{d}{dt}b^-_n+i\bar{a}^{+}_n\frac{d}{dt}a^+_n-i\bar{a}^{-}_n\frac{d}{dt}a^-_n\right)-H^++H^- - i\sum_j  \Gamma_j\left(2L_j^+\bar{L}^-_j-\bar{L}_j^+L_j^+-\bar{L}_j^-L_j^-\right) 
\end{align}
and the Hamiltonian is given by 
\begin{equation}
    H=-\sum_{n=1}^N t(\bar{a}_{n+1}a_n+\bar{a}_{n}a_{n+1})+\mu_n\bar{a}_n a_n+t'(\bar{a}_nb_n+e^{i\phi}\bar{a}_{n+1}b_n+\bar{b}_na_n+e^{-i\phi}\bar{b}_na_{n+1})+M \bar{b}_n b_n
\end{equation}
The dissipation occurs entirely through the upper sites, so the jump operators can be taken to be $L_j=b_j$, and we let $\Gamma_j=\Gamma$ be the same for each site. In order to simplify the notation we define $B_n=(b^+_n,b^-_n)^T$. Then the part of the action involving the $b$ operators is
\begin{equation}
    S_B=\int dt\ \sum_{n=1}^N B^{\dagger}_n D^{-1}B_n+B^{\dagger}_n J_n+J^{\dagger}_n B_n
\end{equation}
where we defined the inverse Green's function
\begin{equation}
    D^{-1}=\begin{bmatrix}
        i\frac{d}{dt}+M+i\Gamma & 0\\
        -2i\Gamma & -i\frac{d}{dt}-M+i\Gamma
    \end{bmatrix}
\end{equation}
and $J_n=(t'(a^+_n+e^{-i\phi}a^+_{n+1}),-t'(a^-_n+e^{-i\phi}a^-_{n+1}))^T$. The action $S_B$ is quadratic, so the integral over the $b$ fields may be carried out. This procedure corresponds to moving from (a) to (b) in figure \ref{fig:sm1}. We obtain the following contribution to the effective action for the $a$ fields
\begin{equation}
    -\sum_{n=1}^{N}J^{\dagger}_n D J_n=\sum_{n=1}^{N}J^{\dagger}_n \begin{bmatrix}
        \frac{1}{i\frac{d}{dt}+M+i\Gamma}&0\\
        -\frac{2i\Gamma}{\Gamma^2+(M+i\frac{d}{dt})^2}&\frac{1}{i\Gamma-i\frac{d}{dt}-M}
    \end{bmatrix} J_n.
\end{equation}
Now we make the assumption that $M$ and $\Gamma$ are large, implying that the timescale of the degrees of freedom that have been integrated out is small compared to the remaining system of interest. Then to leading order
\begin{equation}
    \begin{bmatrix}
        \frac{1}{i\frac{d}{dt}+M+i\Gamma}&0\\
        -\frac{2i\Gamma}{\Gamma^2+(M+i\frac{d}{dt})^2}&\frac{1}{i\Gamma-i\frac{d}{dt}-M}
    \end{bmatrix}\approx \begin{bmatrix}
        \frac{M-i\Gamma}{M^2+\Gamma^2}&0\\
        -\frac{2i\Gamma}{\Gamma^2+M^2}&\frac{-M-i\Gamma}{M^2+\Gamma^2}
    \end{bmatrix}
\end{equation}
so 
\begin{align}
    -\sum_{n=1}^{N}J^{\dagger}_n D J_n\approx (t')^2\sum_{n=1}^N\Bigl[&-\frac{M}{M^2+\Gamma^2}(2\bar{a}^+_n a^+_n+e^{-i\phi}\bar{a}^+_na^+_{n+1}+e^{i\phi}\bar{a}^+_{n+1}a^+_{n})\nonumber\\+&\frac{M}{M^2+\Gamma^2}(2\bar{a}^-_n a^-_n+e^{-i\phi}\bar{a}^-_na^-_{n+1}+e^{i\phi}\bar{a}^-_{n+1}a^-_{n})\nonumber\\&-\frac{i\Gamma}{M^2+\Gamma^2}\left(2\ell_n^{+}\bar{\ell}_n^--\bar{\ell}^+_n\ell^+_n-\bar{\ell}^-_n\ell^-_n\right)\Bigr]
\end{align}
where we defined $\ell_n=a_n+e^{-i\phi}a_{n+1}$. Taking $\phi=\frac{\pi}{2}$ yields the same jump operators as in the main text. Note that the integration led to additional nearest-neighbor hopping terms, which renormalize hopping parameter and the on-site potential in the Hamiltonian. 
Using a Hubbard-Stratonovich transformation it can be shown that the addition of interactions in the $b$ fields does not impact the above calculation at the order we are working to.

\section{2. Details on the Algebraic Bethe Ansatz}
We show this by generalizing the argument in \cite{zheng2023exact}. The R-matrix of this Hamiltonian is the R-matrix of the isotropic Heisenberg model \cite{Korepinbook}
\begin{equation} \label{R-matrix}
    R(\lambda)=\begin{bmatrix}
    f(\lambda)&0&0&0\\
    0&1&g(\lambda)&0\\
    0&g(\lambda)&1&0\\
    0&0&0&f(\lambda)
    \end{bmatrix}
\end{equation}
where 
\begin{equation}
    f(\lambda)=1-\frac{U}{\lambda},\hspace{1cm}g(\lambda)=-\frac{U}{\lambda}.
\end{equation}
Take the Lax operator
\begin{equation}
    L_{j,a}(\lambda)=\begin{bmatrix}
        \lambda-\sqrt{U}n_j &  g a_j^{\dagger}\\
         g a_j& -\frac{g^2}{\sqrt{U}}
    \end{bmatrix},
\end{equation}
which is a function of the rapidity $\lambda$.
The Lax operator acts on $\mathcal{H}_j\otimes \mathcal{H}_{a}$ where $\mathbf{H}_j$ is the Hilbert space of site $j$ in the system -- the "quantum space" -- and $\mathcal{H}_a=\mathcal{C}^2$ is the "auxiliary space". It can be shown to satisfy the $RLL$ equation \cite{Korepinbook}
\begin{equation} \label{RLL}
    R_{12}(\lambda-\nu)L_{j,a_1}(\lambda)L_{j,a_2}(\nu)=L_{j,a_2}(\nu)L_{j,a_1}(\lambda)R_{12}(\lambda-\nu).
\end{equation}
In this equation there are two auxiliary spaces, so that the two Lax operators act on one each and the R-matrix intertwines them. 
The next step is to construct the monodromy matrix
\begin{equation} \label{monodromy}
    T(\lambda,\{\mu_j'\})_a=K L_{1,a}(\lambda-\mu_1')L_{2,a}(\lambda-\mu_2')...L_{N,a}(\lambda-\mu_N'),
\end{equation}
where $K=\mathrm{Diag}(1,\epsilon)$. More explicitly the monodromy matrix is
\begin{equation} \label{monodromy_matrix}
    T(\lambda,\{\mu_j'\})=\begin{bmatrix}
    \mathcal{A}(\lambda,\{\mu_j'\})&\mathcal{B}(\lambda,\{\mu_j'\})\\
    \epsilon\mathcal{C}(\lambda,\{\mu_j'\})&\epsilon\mathcal{D}(\lambda,\{\mu_j'\})
    \end{bmatrix}
\end{equation}
where the elements act on all the quantum spaces. Using Eq.~\eqref{RLL} it can be shown that the monodromy matrix satisfies the $RTT$ relation
\begin{align} \label{RTT}
    R_{12}(\lambda-\nu)&T_1(\lambda,\{\mu_j\})T_2(\nu,\{\mu_j\})=\\\nonumber &T_2(\nu,\{\mu_j\})T_1(\lambda,\{\mu_j\})R_{12}(\lambda-\nu).
\end{align}
In order to get rid of the auxiliary spaces we trace over them, thus defining the transfer matrix 
\begin{equation}
    t(\lambda)=\mathrm{Tr}_a\  T(\lambda,\{\mu_j'\})=\mathcal{A}(\lambda,\{\mu_j'\})+\epsilon\mathcal{D}(\lambda,\{\mu_j'\}),
\end{equation}
which can be expanded in a series
\begin{equation}
    t(\lambda)=\sum_{n=0}^N \lambda^n Q_n.
\end{equation}
Tracing over the auxiliary spaces in the $RTT$ relation we see that $[t(\lambda),t(\nu)]=0,$ which in turn implies that all $Q_n$ are in involution, $[Q_n,Q_m]=0$. Integrable Hamiltonians are constructed by combining the $Q_n$, and in particular the effective non-Hermitian Hamiltonian discussed in the main text is obtained by the combination \cite{zheng2023exact}
\begin{equation} \label{nh_hamil_comb}
    H_{eff}=\frac{1}{2}\left(Q_{N-1}^2+\sqrt{U}Q_{N-1}\right)-Q_{N-2},
\end{equation}
if we identify $g^2=2t$. In order to write down the eigenvectors we need a vacuum $\ket{\Omega}$ which satisfies
\begin{align}
    \mathcal{A}(\lambda)\ket{\Omega}&=a(\lambda)\ket{\Omega}\nonumber\\
    \mathcal{D}(\lambda)\ket{\Omega}&=d(\lambda)\ket{\Omega}\nonumber\\
    \mathcal{C}(\lambda)\ket{\Omega}&=0
\end{align}
for c-numbers $a(\lambda),d(\lambda),$ which are called the vacuum eigenvalues. In the present model $a(\lambda)=\prod_{j=1}^N (\lambda-\mu_j), d(\lambda)=\epsilon(-2t/U)^N$. The states given by the algebraic Bethe Ansatz
\begin{equation}
    \ket{\{\lambda_n\}}=\prod_{n=1}^M\mathcal{B}(\lambda_n)\ket{\Omega},
\end{equation}
can be shown to be eigenstates of $t(\lambda)$ if the rapidities $\{\lambda_k\}$ satisfy the Bethe equations 
\begin{equation} \label{gen_bethe_eqs}
    a(\lambda_n)\prod_{j\neq n}(\lambda_n-\lambda_j-U)-d(\lambda_n)\prod_{j\neq n}(\lambda_j-\lambda_n-U)=0.
\end{equation}
In order to prove this, one requires that
\begin{equation}
    t(\lambda) \ket{\{\lambda_n\}}=\Lambda(\lambda,\{\lambda_n\})\ket{\{\lambda_n\}} \iff (\mathcal{A}(\lambda)+\epsilon\mathcal{D}(\lambda))\prod_{n=1}^M\mathcal{B}(\lambda_n)\ket{\Omega}=\Lambda(\lambda,\{\lambda_n\})\prod_{n=1}^M\mathcal{B}(\lambda_n)\ket{\Omega}
\end{equation}
for some eigenvalue $\Lambda(\lambda,\{\lambda_n\})$. The $\mathcal{A}$ and $\mathcal{D}$ operators may be commuted through all the $\mathcal{B}$ operators using the $RTT$ relation \eqref{RTT}. This leads to the Bethe equations, and to an expressions of the eigenvalues: 
\begin{equation}
    \Lambda(\lambda,\{\lambda_n\})=a(\lambda)\prod_{n=1}^Mf(\lambda-\lambda_n)+d(\lambda)\prod_{n=1}^Mf(\lambda_n-\lambda).
\end{equation}
The Bethe equations can also be obtained by requiring that the residue of $\Lambda$ is $0$ at each pole. Using this expression the eigenvalues of the operators $Q_i$ may be deduced by performing a series expansion in $\lambda$ and inserting the Bethe equations. 
\section{3. Diagonalizing the Liouvillian}
In the main text the following Liouvillian is discussed
\begin{equation} 
    \frac{d\rho}{dt}=\mathcal{L}\rho=-i[H,\rho]+\sum_j  \gamma_j\left(2\ell_j\rho\ell^{\dagger}_j-\{\ell_j\ell_j^{\dagger},\rho\}\right).
\end{equation}
where $H$ is the disordered 1d Bose-Hubbard Hamiltonian
\begin{align} 
    H=-&t\sum_{j=1}^N a_j^{\dagger}a_{j+1}+a_{j+1}^{\dagger}a_j+\frac{U}{2}\sum_{j=1}^N n_j(n_j-1)-\sum_{j=1}^N  \mu_j n_j,\nonumber\\
    &-\frac{1}{2}\epsilon (a^{\dagger}_{N}a_1+a^{\dagger}_1a_N)
\end{align}
and $\ell_j=a_j+i a_{j+1}$ for bosonic creation and annihilation operators $a^{\dagger}_j,a_j$. It is triangular in the eigenbasis of the non-hermitian effective Hamiltonian, which for the choice $\gamma_j=t$ is given by for $j\neq N$ and $\gamma_N=\frac{\epsilon}{2}$ is given by
\begin{align} 
    H_{eff}&=-2t\sum_{j=1}^N  a_{j+1}^{\dagger}a_j+\frac{U}{2}\sum_{j=1}^N n_j(n_j-1)-\sum_{j=1}^N  \mu_j'n_j\nonumber\\
    &-\epsilon a^{\dagger}_1 a_N.
\end{align} 
Since the Liouvillian is triangular its eigenvalues the same as its diagonal elements, which are identical to the differences between eigenvalues of the non-hermitian effective Hamiltonian. These eigenvalues have the form
\begin{equation}
    \omega(\{\lambda_n\}_{n=1}^M,\{\nu_m\}_{m=1}^{M'})=-i\sum_{n=1}^M\lambda_n+i\sum_{m=1}^{M'}\bar{\nu}_m
\end{equation}
where $\{\lambda_n\}_{n=1}^{M},\{\nu_m\}_{m=1}^{M'}$ are two solutions to the Bethe equations and a overhead bar indicates complex conjugation.
The explicit form of the eigenoperators of a Liouvillian of this kind was given in Ref. \onlinecite{Torres2014} in terms of correlation functions and eigenvalues of the effective non-Hermitian Hamiltonian. For the model under consideration here the eigenoperators may be written down in terms of the solutions of the Bethe equations, as discussed in the main text. Following the notation of Ref. \onlinecite{Torres2014} we define the matrix
\begin{align}
    \mathcal{F}^{(l,n)}_{(\lambda;\nu),(\lambda';\nu')}=&\sum_{j=1}^N\prescript{}{L}{\bra{\{\lambda_k\}_{k=1}^{n-1}}}\ell_j\ket{\{\lambda'_n\}_{k=1}^{n}}_R\nonumber\\
    &\times \prescript{}{L}{\bra{\{\nu_l\}_{l=1}^{n+l-1}}}\ell_j\ket{\{\lambda'_l\}_{l=1}^{n+l}}_R,
\end{align}
where the $R$ and $L$ subscripts indicate right and left eigenstates respectively, and the diagonal matrix
\begin{equation}
    T^{(l,m;n)}_{(\lambda;\nu),(\lambda';\nu'),(\lambda'';\nu'')}=\left(\omega_{\lambda,\nu}^{(l,m)}-\omega_{\lambda',\nu'}^{(l,n)}\right)^{-1}\delta_{(\lambda',\nu'),(\lambda'',\nu'')}
\end{equation}
where $\omega_{\lambda,\nu}^{l,m}=\omega(\{\lambda_i\}_{i=1}^{m+l},\{\nu_j\}_{j=1}^{m})$.  Next we define the column vectors
\begin{align}
    v_{(\lambda,\nu)}^{l,m;n}=\left(\prod_{i=n}^{m-1}T^{l,m;i}_{(\lambda,\nu)}\mathcal{F}^{(m,i+1)}\right)e^{(l,m)}_{(\lambda,\nu)}.
\end{align}
The product here is a matrix product, and $e^{(l,m)}_{(\lambda,\nu)}$ is a column vector which is $1$ at entry $(\lambda,\nu)$ and $0$ everywhere else. Using these definitions we can write the right eigenoperators of the Liouvillian as
\begin{equation} 
    \rho_{(\lambda,\nu)}=\sum_{n=0}^m\sum_{(\lambda',\nu')}v^{(l,m;n)}_{(\lambda,\nu);(\lambda',\nu')}\ket{\{\lambda'\}}\bra{\{\nu'\}},
\end{equation}
and the left eigenvectors may be calculated in a similar way. The second sum is over all solutions of the Bethe equations with particle number less than or equal to $n$. 

 Recalling that the jump operators are $\ell_j=a_j+ia_{j+1}$, we see that we need to calculate form factors like $\prescript{}{L}{\bra{\{\lambda\}}}a_k\ket{\{\nu\}}_R$. In principle these may be computed by brute force, since the eigenstates are known. However, if the system is translationally invariant, so that $\mu_n=\mu$ and $\epsilon=t$, then they can be written explicitly in terms of the solutions of the Bethe equations. First, we use the biorthogonal completeness relation \cite{Brody2014} to write
\begin{equation} \label{LR_form_factors}
    \prescript{}{L}{\bra{\{\lambda\}}}a_k\ket{\{\nu\}}_R=\sum_{\{\lambda'\}}\prescript{}{R}{\bra{\{\lambda\}}}a_k\ket{\{\lambda'\}}_R\prescript{}{L}{\braket{\{\lambda'\}|\{\nu\}}_L}
\end{equation}
where the $R,L$ subscripts indicate if a state is a right eigenstate or a left eigenstate. The scalar products may be calculated using Slavnov's determinant representation \cite{Slavnov1989,Slavnovbook} if the two states are not the same, and using Gaudin's formula for norms of Bethe states \cite{Gaudin} if they are. Alternatively more general expression for scalar products in integrable models may be used \cite{Korepinbook}.

The other factors that appear in this sum -- the right-right form factors $\prescript{}{R}{\bra{\{\lambda\}}}a_k\ket{\{\lambda'\}}_R$ -- are typically challenging to calculate in integrable models, but in the translationally invariant case a completely explicit expression for this type of correlation function for any integrable model with the XXX-model R-matrix was given in Ref. \onlinecite{Oota2004}. 
Then the local operator $a_k$ may be written as
\begin{equation} \label{local_operator}
    a_k= \mathcal{U}^{k-1}\mathcal{D}^{-1}(0)\mathcal{C}(0)\mathcal{U}^{-k+1}
\end{equation}
where $\mathcal{U}$ is the shift operator. It can be shown that the Bethe eigenvector $\ket{\{\lambda_n\}}$ is also an eigenvector of $\mathcal{U}$ with eigenvalue $\prod_{n}\frac{-\lambda_n}{2t}$ \cite{Korepinbook}. Hence
\begin{align}
    \prescript{}{R}{\bra{\{\lambda\}}}a_k\ket{\{\lambda'\}}_R&=\prescript{}{R}{\bra{\{\lambda\}}}\mathcal{U}^{k-1}\mathcal{D}^{-1}(0)\mathcal{C}(0)\mathcal{U}^{-k+1}\ket{\{\lambda'\}}_R\nonumber\\
    &=-\prod_{n=1}^M\lambda_n^{k-1}\prod_{n=1}^{M+1}(\lambda'_n)^{-k+1}\prescript{}{R}{\bra{\{\lambda\}}}\mathcal{D}^{-1}(0)\mathcal{C}(0)\ket{\{\lambda'\}}_R.
\end{align}
The remaining unknown factor in this expression has a determinant representation. Following Ref. \cite{Oota2004} we define
\begin{equation}
    t(\lambda)=\frac{1}{\lambda(\lambda-1)}
\end{equation}
and the matrix $S$ with elements
\begin{equation}
    S_{kl}=t(\lambda'_k-\lambda_l)\frac{\prod_{m=1}^M(1+\lambda_m-\lambda'_l)}{\prod_{m=1}^{M+1}(1+\lambda'_m-\lambda'_l)}-t(\lambda_k-\lambda'_l)\frac{\prod_{m=1}^M(1+\lambda'_l-\lambda_m)}{\prod_{m=1}^{M+1}(1+\lambda'_l-\lambda'_m)}.
\end{equation}
Using these definitions one can show that
\begin{align}
    \prescript{}{R}{\bra{\{\lambda\}}}\mathcal{D}^{-1}(0)\mathcal{C}(0)\ket{\{\lambda'\}}_R= &(-1)^{M} (2t)^{2NM+M}\left(\det S\right) \prod_{k,l=1}^{M+1}(1+\lambda'_k-\lambda'_l) \nonumber\\
    &\times \prod_{1\leq k<l\leq M} \frac{1}{(\lambda_l-\lambda_k)(\lambda'_l-\lambda'_k)} \prod_{k=1}^M\frac{1}{\lambda'_{M+1}-\lambda_l}\frac{\prod_{l=1}^{M}\lambda_l}{\prod_{l=1}^{M+1}\lambda'_l}
\end{align}
This reduces the calculation of eigenoperators of the Lindbladian without disorder to solving the Bethe equations \eqref{gen_bethe_eqs}. However, these methods rely heavily on translational invariance, so in the case of open boundary conditions, or if disorder is present, calculating the elements of the matrix $\mathcal{F}^{(l,n)}$ cannot obviously be done by any other method than brute force calculations using the coordinate Bethe Ansatz.
\FloatBarrier
\section{4. The particle densities -- numerical results}
In this section we present some numerical calculations of the particle densities for the unidirectional disordered Bose-Hubbard model with open boundary conditions that is considered in the main text. The results in the figures were obtained using exact diagonalization. 

\begin{figure}
    \centering
    \includegraphics[width=0.7\textwidth]{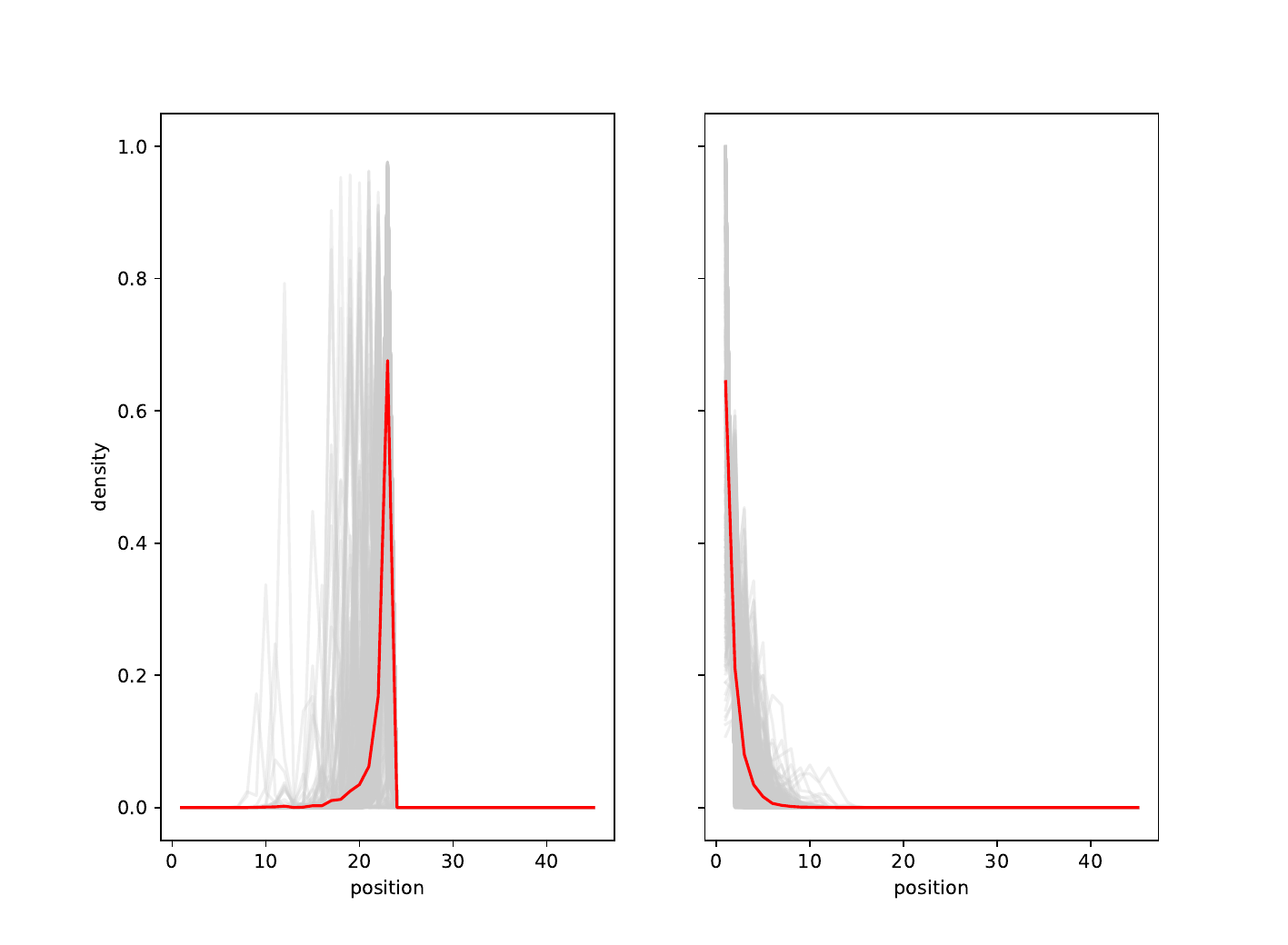}
    \caption{ Density of single particle ground states with $\Delta=1$ and $N=45$. The minimum of the potential was chosen to be at position $22$, and the rest were chosen to be uniformly distributed in the interval $[-\Delta,\Delta]$ with $500$ disorder iterations. The grey lines correspond to individual disorder realizations, while the red line is the average. In the left figure the hopping parameter is $t=0.15$ while in the right figure the hopping parameter is $t=0.8$.}
    \label{fig:num_noninteracting}
\end{figure}

Figure \ref{fig:num_noninteracting} shows the disorder-averaged density $ \bra{GS}a^{\dagger}_x a_x\ket{GS}/\langle GS|GS\rangle$ of the non-interacting ground state. The disorder was chosen so that its minimum $-\Delta=-1$ was located at position $22$ and the rest was chosen to be uniformly distributed in the interval $[-\Delta.\Delta]$. Note that the number of particles in the system was chosen to be $1$, since many-body effects are not relevant in the non-interacting case. We see that for $t=0.15\Delta$ the particle is localized at the minimum of the potential, which is located at site $22$, while for $t=0.8\Delta$ the particle is localized at the boundary. There must therefore be a transition between these phases when $t$ is varied in the interval $[0.15\Delta,0.8\Delta]$, in agreement with the main text, where it was shown that this transition must occur for some $t_c\leq \Delta$.

\begin{figure}
    \centering
    \includegraphics[width=0.7\textwidth]{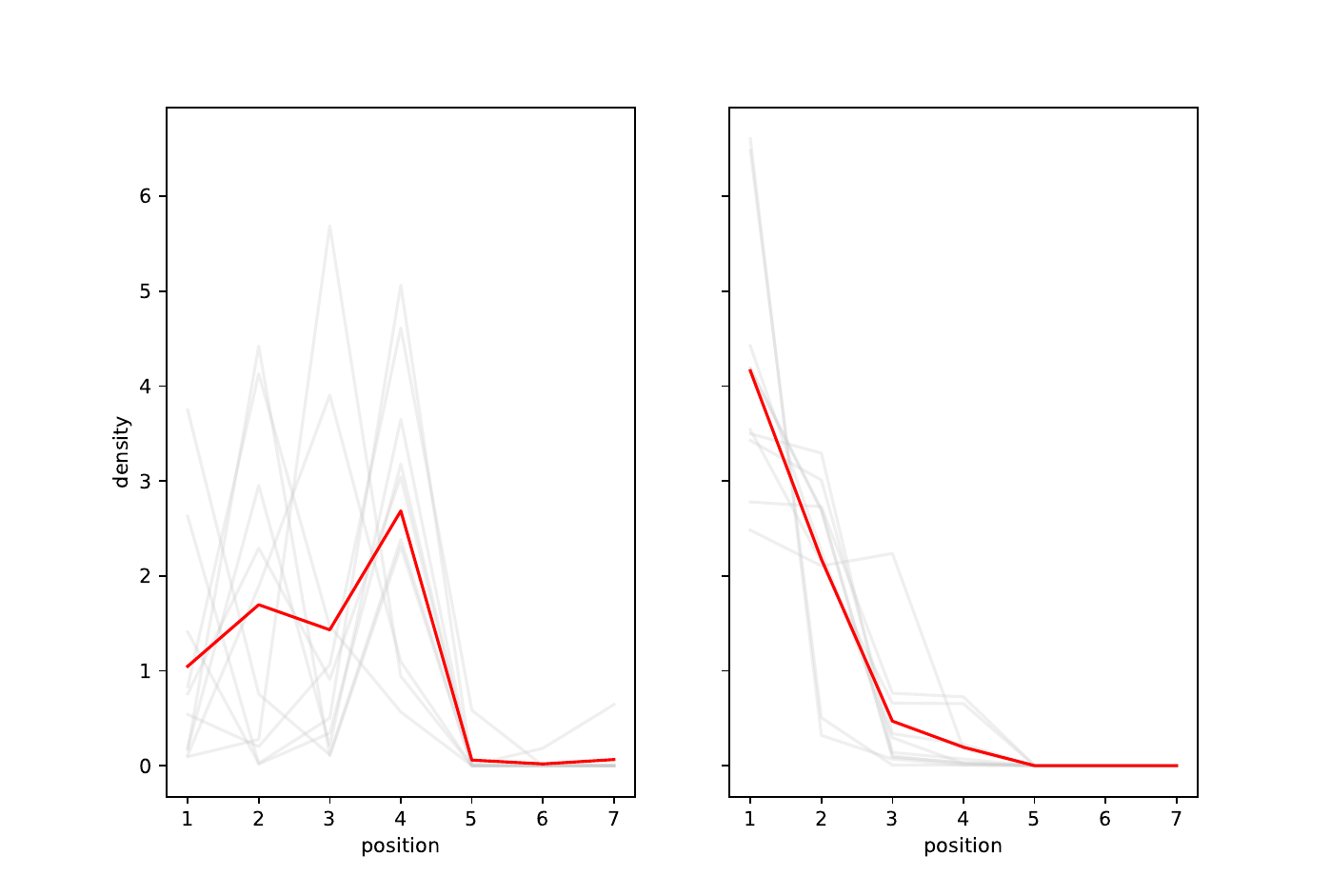}
    \caption{ Density of the many-particle ground states with $\Delta=1$, $U=0.1$ and unit filling, $N=M=7$. The minimum of the potential was chosen to be at position $22$ and the rest were chosen to be uniformly distributed in the interval $[-\Delta,\Delta]$ with $10$ disorder iterations The grey lines correspond to individual disorder realizations, while the red line is the average. In the left figure the hopping parameter is $t=0.15$ while in the right figure the hopping parameter is $t=0.8$.}
    \label{fig:num_bose_glass}
\end{figure}

Figure \ref{fig:num_bose_glass} again shows the disorder-averaged density, but now in the interacting case. The disorder was chosen so that its minimum $-\Delta=-1$ was located at position $3$ and the rest was chosen to be uniformly distributed in the interval $[-\Delta.\Delta]$. The interaction strength $U$ was chosen to be small, $U=0.01$, to ensure that the system is in the fragmented condensates phase discussed in the main text. We see that for $t=0.15\Delta$ the largest fraction of particles are localized at $3$ while for $t=0.8\Delta$ all the particles are localized at the boundary, so there must be a phase transition when $t$ is varied in the interval $[0.15\Delta,0.8\Delta]$, which is in agreement with the main text, since $U$ is small.

\begin{figure}
    \centering
    \includegraphics[width=0.7\textwidth]{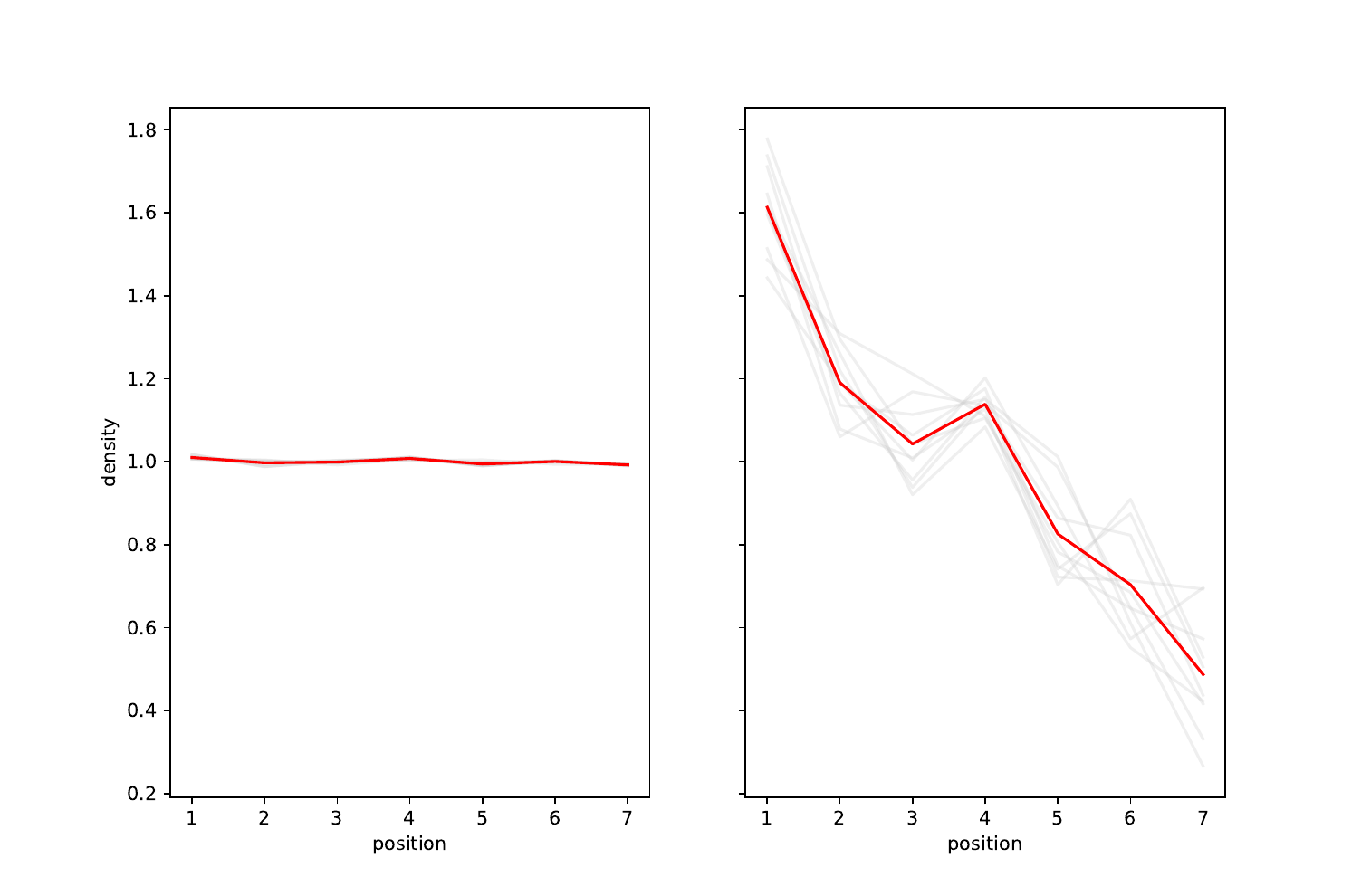}
    \caption{ Density of the many-particle ground states with $\Delta=1$, $U=5$ and unit filling, $N=M=7$ . The minimum of the potential was chosen to be at position $22$ and the rest were chosen to be uniformly distributed in the interval $[-\Delta,\Delta]$ with $10$ disorder iterations The grey lines correspond to individual disorder realizations, while the red line is the average. In the left figure the hopping parameter is $t=0.15$ while in the right figure the hopping parameter is $t=0.8$.}
    \label{fig:num_mott}
\end{figure}

Finally, figure \ref{fig:num_mott} shows the disorder-averaged density in the Mott phases described in the main text. Again, the disorder was chosen so that its minimum $-\Delta=-1$ was located at position $3$ and the rest was chosen to be uniformly distributed in the interval $[-\Delta.\Delta]$. However, now the interaction strength was chosen to be large, $U=5$, to ensure that the system is in one of the Mott phases. We see that for $t=0.15\Delta$ the particles are evenly distributed across the lattice, while for $t=0.8\Delta$ the particles are localized at the boundary. In the latter case they are not as sharply localized as in the fragmented condensates case. Nonetheless, since the system is at unit filling this is in agreement with the main text, since there must be a transition as $t$ is varied in the interval $[0.15\Delta,0.8\Delta]$.

\end{document}